\documentstyle [prl,aps,multicol,amstex,epsfig]{revtex}
\begin{document}
\draft
\hyphenation{following}
\title{Time-reversal symmetry in nonlinear optics}
\author{M. Trzeciecki$^{1,2}$ and W. H\"ubner$^1$}
\address{$^1$Max-Planck-Institut f\"ur Mikrostrukturphysik, Weinberg 2,
D-06120 Halle, Germany}
\address{$^2$Institute of Physics, Warsaw University of Technology, Koszykowa 
75, 00-662 Warsaw, Poland}
\date{\today}
\maketitle

\begin{abstract}
The applicability of time-reversal symmetry to nonlinear optics is 
discussed, both from macroscopic (Maxwell equations) and microscopic (quantum 
theoretical) point of view. We find that only spatial operations can be applied 
for the symmetry classification of nonlinear optical processes in 
magnetic, in particular antiferromagnetic, materials. An example is given where 
both operations (time reversal and a spatial operation) can yield different 
results.
\end{abstract}
\begin{multicols} {2}

Symmetries determine several important properties of a crystal, in 
particular its optical response. In magnetic materials,
time-reversal is believed to be of fundamental importance since this operation 
reverses all magnetic moments \cite{ref4,ref13,ref17}.
However, the consequences of applying time-reversal are more 
profound than a simple inversion of localized magnetic moments. As it will turn 
out in this paper there is a deep interrelation between the absence of 
conventional dissipation in even-order (e.g. second) harmonic generation and 
the influence of time-reversal on spin ordering. 
This brings about a subtle difference between time-reversal and spatial 
symmetries in 
nonlinear optics. The benefit of this difference makes optical second harmonic 
generation (SHG) a rather unique probe of antiferromagnetism, while linear 
optics (where dissipation in the conventional sense is possible) is blind for 
such balanced spin structures. The recent 
discussion about the influence of micro-irreversibility on macro-reversibility 
and reciprocity (\cite{ref187,ref188,ref190}) shows that the issue of 
time-reversal, although extensively discussed, is far from being understood.

The theory of nonlinear optics has been developed since the 60s. The 
pioneering work of Armstrong {\em et al.}~\cite{ref185} describes the 
propagation of a light wave through a nonlinear medium, where the energy may be 
converted from the fundamental frequency to higher harmonics (or vice versa). An 
exhaustive description of nonlinear optical phenomena is contained in the 
fundamental books by Bloembergen \cite{blombook} and Shen \cite{shenbook}. In 
these works (\cite{ref185,blombook,shenbook}), a unique flow of time is tacitly 
assumed, while {\em magnetism} is entirely absent. Consequently the issue of 
time reversal is not essential for these authors. The discussion of magnetism 
has been brought to nonlinear optics by Pan {\em et al.} \cite{ref53} and 
H\"ubner {\em et al.} \cite{ref80}. In these papers, time-reversal was applied 
to reverse the localized magnetic moments, since the discussion was focused on 
ferromagnetism. However, the experimental observation of antiferromagnetic (AF) 
domains in Cr$_2$O$_3$ by Fiebig {\em et al.} \cite{ref18} and the subsequent 
theoretical analyses by Muthukumar {\em et al.} \cite{ref27} and by D\"ahn {\em 
et al.} \cite{dahn} challenged the validity of time-reversal for the symmetry 
analysis of optical processes. Since the inclusion or absence of time-reversal 
in the theoretical analysis of SHG from antiferromagnets yields different 
predictions of the experimental results, the issue is shifted from 
academic interest to practical relevance. The importance of the theoretical 
analysis of SHG from antiferromagnets is tremendously growing due to the unique 
capabilities of this method in probing buried AF layers, which in turn is 
important for the characterization of recently upcoming magnetoelectronic 
devices such as tunneling magnetoresistive junctions.

In considering the time-reversibility of an experimental situation, three 
approaches are possible: (i) time-reversal is applied to the sample, but all the 
processes resulting from the experiment are unchanged. In particular, the 
magnetic moments in the 
sample are reversed, but the direction of the light propagation through the 
sample is not affected. This approach is presented e.g. in  
\cite{ref166,ref189}. We consider this approach as incomplete, since it does 
not equally treat the sample and the light propagating through it. (ii) The 
second approach, usually encountered in the so-called Sagnac-interferometry, 
addresses time-reversal by reversing the propagation of the light through the 
sample (see, e.g. \cite{ref154,ref155,ref156}). Clearly, such procedure probes 
the {\em reciprocity} of the sample rather than its time-reversal symmetry. It 
can also be proven that the second approach is equivalent to the first one. 
(iii) According to the third approach, presented e.g. 
in \cite{Shelankov}, time reversal acts on {\em both}: the sample and the 
experimental setup. In this letter, we will follow approach (iii).

In the processes of even-order harmonic generation, dissipation in the 
conventional sense, converting radiation into heat, does not exist, since the 
energy loss of the electromagnetic field is the time average \cite{Bassani}
\begin{equation}
\label{eqDiss}
-\biggl \langle {dP(t)\over dt}E(t) \biggr \rangle,
\end{equation}
which vanishes for SHG (and all even-order harmonics), since 
\begin{equation}
\label{cancel}
\begin{split}
P(t) &\sim P_0 e^{i\omega t} \\
E(t) &\sim E_0 e^{i2\omega t}
\end{split}
\end{equation}
Here, $P$ and $E$ denote the polarization of the medium and the electric field, 
respectively \cite{attenuation}. The lack of dissipation in the conventional 
sense does not mean that the process of SHG is reversible. Already the analysis 
by Armstrong {\em et al.} \cite{ref185} assumes a unique time 
direction. There, the nonlinear polarization ${\mathbf{P}}^{NL}$ 
and the electric field {\bf E}$_3$ of a light beam resulting from Sum Frequency 
Generation at a point $r_0$ is given by:
\begin{equation}
\label{bl3_1}
	{\mathbf{P}}^{NL}(\omega_3) \sim \frac{1}{2}{\text{Re}} \biggl [
	e^{i({\mathbf{\Delta k \cdot r_0}}+\Delta \phi)} e^{i({\mathbf{k_3 r_0}} 	
	- \omega_3 t + \phi_3)}
	\biggr ]
\end{equation}
\begin{equation}
\label{bl3_2}
	{\mathbf{E}}_3 \sim {\text{Re}} \biggl [ e^{i({\mathbf{k_3 \cdot r_0}} - 	
	\omega_3t + \phi_3)} \biggr ],
\end{equation}
see eqs. (3.1) and (3.2) of Ref. \cite{ref185}.
Here, $\omega_3$ and $k_3$ describe the frequency and wave vector of the 
generated light ($\omega_3=\omega_1+\omega_2$ and ${\mathbf{k_3}} \approx 
{\mathbf{k_1}}+{\mathbf{k_2}}$). The authors introduce the idea of ``work done 
on this wave'' by the nonlinear polarization of the medium, equal to 
\begin{multline}
\label{bl3_4}
W_3=\frac{\omega_3}{2\pi} \int_{cycle} {\mathbf{E}}_3 
\frac{d{\mathbf{P}}^{NL}(\omega_3)}{dt} dt \\
= \frac{1}{2} \omega_3 {\mathbf{E}}_3 {\mathbf{P}}^{NL}(\omega_3, 
\text{out-of-phase}),
\end{multline}
if the polarization is exactly 90$^\circ$ out of phase with the electric field 
(which requires that $\Delta k_zz +\Delta \phi=\pi /2$).
The work done on the generated wave {\em determines the direction of time}. This 
presents a new kind of dissipation, namely ``dissipation in the frequency 
space'', which invalidates time-reversal symmetry.

\begin{figure}
\psfig{file=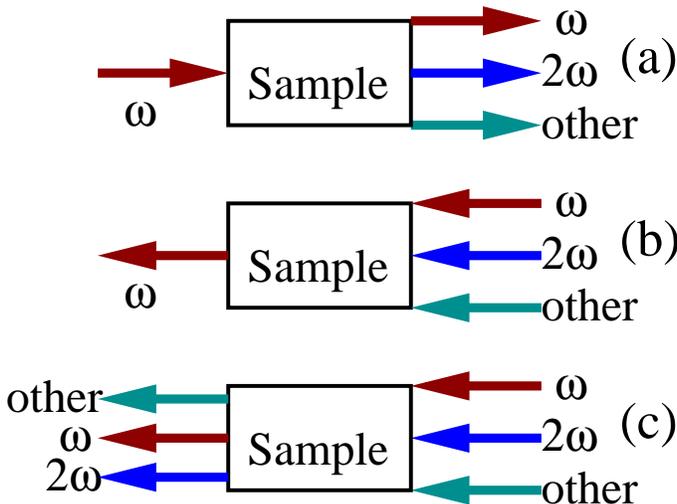, angle=270, width=0.5\linewidth}
\caption{\label{figreversal} Time-reversal asymmetry in SHG. Panel (a) presents 
the original process, panel (b) a process in reversed time which would restore 
the symmetry, panel (c) presents a physically valid process described in 
reversed time.}
\end{figure}

This fact becomes even more obvious if one takes the global picture of SHG. 
Radiation acting on an {\em ensemble of atoms} may excite and deexcite them in 
many ways {\em simultaneously}. Thus contributions of many frequencies are 
always present (see Fig. \ref{figreversal}(a)). One has a unique source of 
$\omega$ light but several detectors for beams of different frequencies: 
$2\omega$, $3\omega$, etc, resulting from sum frequency generation (in 
particular SHG); linearly propagating $\omega$ light; and a DC current resulting 
from difference frequency generation. This is due to the expansion of the source 
term (polarization $\Bbb{P}$) in terms of the electric field:
\begin{multline}
	\Bbb{P}={\mathbf{P_1}}+{\mathbf{P_2}}+\ldots = \\
	=\chi^{(1)}(\omega){\mathbf{E}}^{(\omega)}+
	\chi^{(2)}(\omega){\mathbf{:E}}^{(\omega)}{\mathbf{E}}^{(\omega)}+\ldots
\end{multline}
Imposing time reversal, the detectors become sources and vice versa. Thus, in 
the time reversed process, one ends up with a single detector, the one which 
receives the light of frequency $\omega$ (Fig. \ref{figreversal}(b)). In order 
to obtain this single frequency one has to redirect all these (previously 
generated) beams back to the sample, conserving their phases. The source term 
now becomes:
\begin{multline}
	{\Bbb{P}}=\chi^{(1)}(\omega){\mathbf{E}}^{(\omega)}
	+\chi^{(1)}(2\omega){\mathbf{E}}^{(2\omega)}+\ldots+\\
	+\chi^{(2)}(\omega)\mathbf{:E}^{(\omega)}\mathbf{E}^{(\omega)}
	+\chi^{(2)}(2\omega){\mathbf{:E}}^{(2\omega)}{\mathbf{E}}^{(2\omega)}
	+\ldots	
\end{multline}
Since the phases of the now incident electric fields are the same as for the 
previously outgoing electric fields, all the terms but those with 
$\chi^{(1)}$ cancel (which means that in the outgoing light one now has only the 
contribution at the frequency $\omega$) and the original situation 
at the input of the process is restored. This description, though mathematically 
correct, is physically invalid, since there is no practical way to detect an 
infinite array of frequencies along with the phases and to revert it with  
arbitrary accuracy (Fig. \ref{figreversal}(c)). Tracing out the ``bath'' degrees 
of freedom (frequencies other than $\omega$ and 2$\omega$) causes a transition 
from a pure to a mixed state of the system, which means that some memory is 
lost. This happens because the traced subsystem and the bath are not 
statistically independent \cite{brenig}. Thus, in any practical situation, there 
is no possibility to generate only the frequency $\omega$ out of a whole array 
of frequencies. The process of SHG looks different in (-t) than in (t). Such a 
process is called {\em dynamical}. 

\begin{figure}
\psfig{file=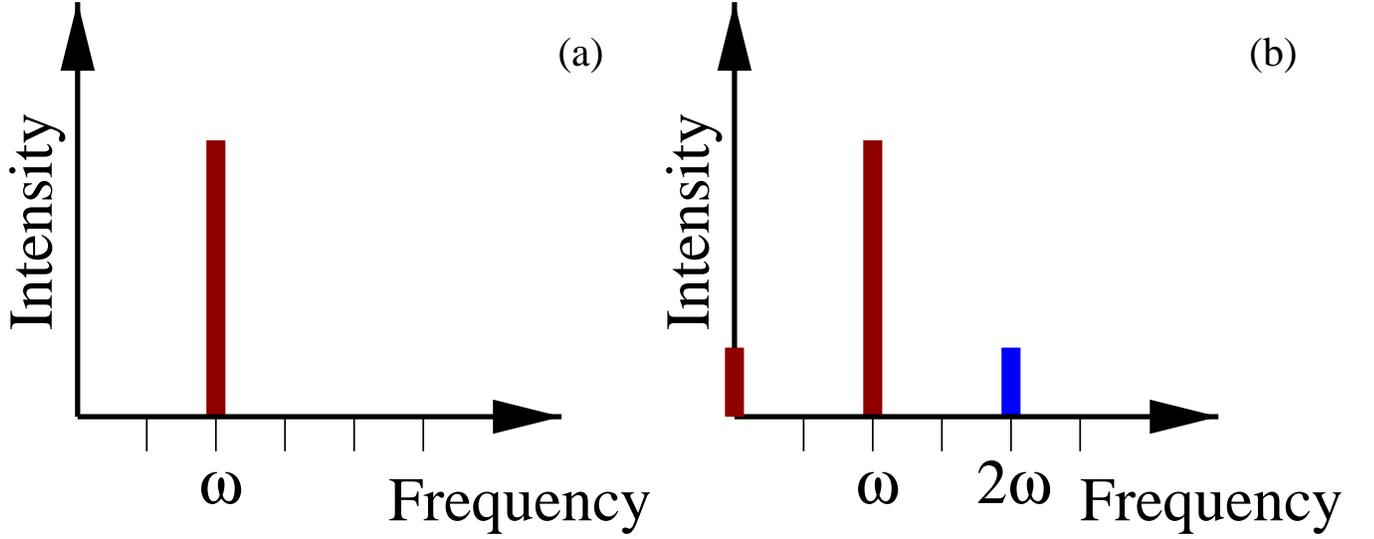, angle=270, width=\linewidth}
\caption{\label{figinten} Light intensity distribution on the input (a) and on 
the output (b) of the SHG.}
\end{figure}

As stated before, there is no dissipation in the process of SHG in the usual 
meaning, i.e. the amount of energy in the radiative form is constant. However, 
there is a transfer of energy between the frequencies, in particular energy 
flows from the frequency $\omega$ to other frequencies (see Fig. 
\ref{figinten}). We  call this {\em dissipation in frequency space}, in contrast 
to the more usual {\em dissipation in real time}. Dissipation in frequency space 
can mix real and imaginary parts of the nonlinear 
susceptibility tensor. The distinction between these two types of dissipation is 
often encountered in the literature. We consider them here on an equal footing 
stating that the presence of any of them (in our case it is the dissipation in 
frequency space) causes the system to have dynamical and thus irreversible 
properties. In this case, time-reversal does not apply to the symmetry analysis 
\cite{ref27,dahn,birss}.

So far we have reasoned that the time-reversal operation has to be excluded from 
the symmetry analysis of SHG. However, {\em magnetism} may bring an additional 
complication, since the magnetic spin structure is an additional aspect the 
symmetry analysis must account for, and it is the time-reversal which is 
conveniently applied to flip the local magnetic moments. This is, however, not 
correct: it is the classical covering symmetry \cite{ref15} of the magnetic 
crystal which should be addressed in a symmetry analysis rather than the 
quantum-mechanical symmetry of the wavefunctions \cite{quantum}. This means that 
the operation applied to reverse the localized magnetic moments should be 
performed in real space rather than Hilbert spin space. Consequently, 
time-reversal cannot be used for the symmetry classification of magnetic 
moments.

Taking into account that time-reversal is not suitable for the 
description of dynamical phenomena, one needs an operation which merely flips 
the localized magnetic moments without inverting the time-flow. This can be 
accomplished by purely spatial 
point-group operations. In many {\em antiferromagnetic} crystals a simple 
translation by a lattice vector reverses the magnetic moments. In many 
ferromagnetic and antiferromagnetic systems this may be accomplished by a mirror 
operation. The spatial operation, which reverses the localized magnetic moments, 
is called by us ``moment-reversal''. This operation is obviously unitary, in 
contrast to the time-reversal operation. Consequently, one does not need to 
invoke the time-reversal operation to describe the full symmetry of magnetic 
crystals. 

\begin{figure}
\psfig{file=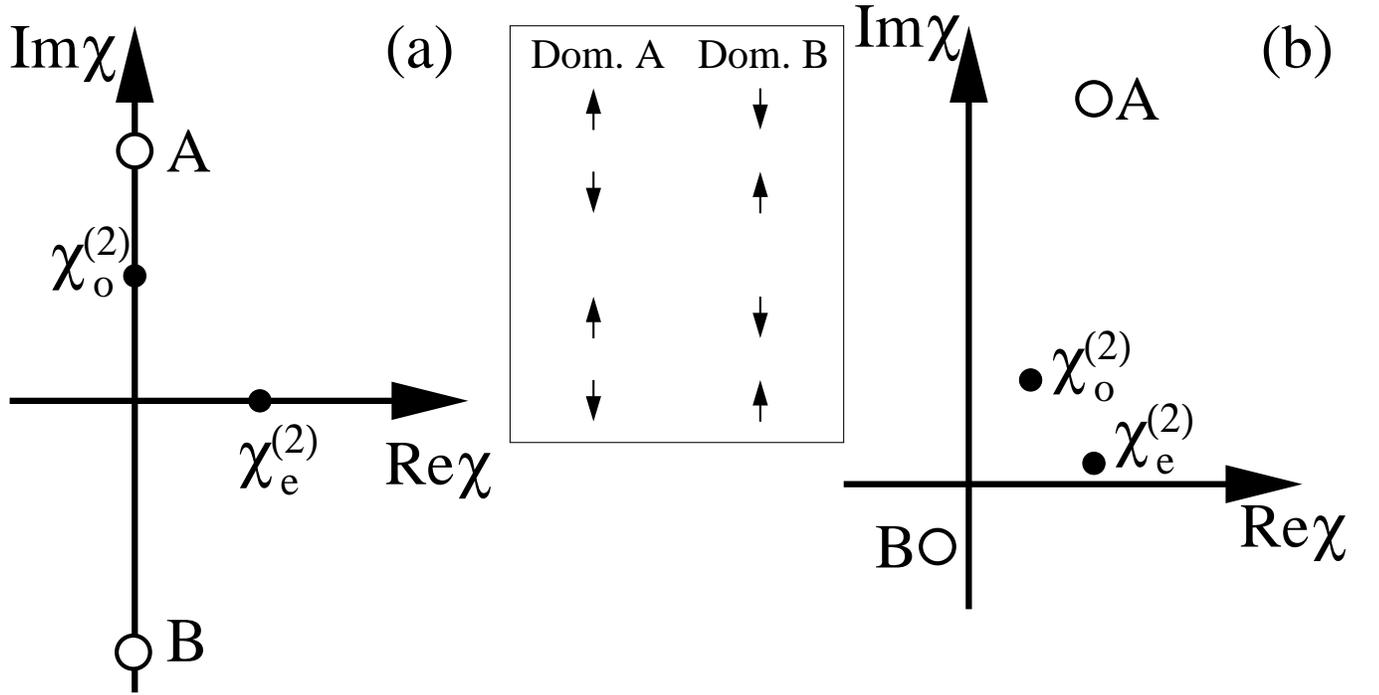, angle=270, width=\linewidth}
\caption{\label{chi} Nonlinear susceptibility tensor elements and resulting SHG 
intensity using time-reversal (panel (a)) and spin-reversal (panel (b)). 
Position of the points ``A'' and ``B'' is given by 
$(\chi^{(2)}_e)^2+(\chi^{(2)}_o)^2\pm2\chi^{(2)}_e \cdot \chi^{(2)}_o$, and the 
distance of the points ``A'' and ``B'' from the origin of the complex plane 
corresponds to the intensity of SHG from the domains A and B, respectively (see 
inset for an example of domains in Cr$_2$O$_3$). For simplicity, the moduli of 
the tensor elements have been taken as equal to 1, but the argumentation also 
holds in the general case.}
\end{figure}

Next, we support our reasoning by an example where the application of 
time-reversal and ``moment-reversal'' in the symmetry analysis yields different 
results (see Fig. \ref{chi}). Let us assume a spin structure with two domains, A 
and B, related to each other by spin-reversal \cite{domains}. A symmetry 
analysis, similar to the one in \cite{Artyk1}, provides us with the set of 
nonvanishing elements of the nonlinear susceptibility tensor (i.e. $\chi^{(2)}$ 
tensor) along with the parities of these elements. Let us assume that for a 
certain experimental geometry only two tensor elements, called $\chi^{(2)}_o$ 
and $\chi^{(2)}_e$, contribute to the resulting SHG light, and that 
$\chi^{(2)}_o$ is odd while $\chi^{(2)}_e$ is even in the domain operation. 
The intensity of SHG light at a fixed polarization is given by:
\begin{equation}
\label{imaging}
I_p\sim |(\chi^{(2)}_e)^2+(\chi^{(2)}_o)^2\pm2\chi^{(2)}_e \cdot \chi^{(2)}_o|
\end{equation}
where ``+'' stands for domain A, ``-'' for domain B. In the conventional 
approach, where {\em time-reversal} is the operation mapping domains into each 
other, $\chi^{(2)}_o$ must be purely imaginary and $\chi^{(2)}_e$ purely 
real (Fig. \ref{chi}(a)), since the operation of time-reversal is anti-unitary 
\cite{tinkham}. In this traditional approach, the first two components 
of the sum in eq. (\ref{imaging}) are real, while the last one is imaginary. 
Because it is the modulus of the whole sum that determines the output intensity, 
the domain contrast is lost since
\begin{equation}
\label{complex}
|\text{a+ib}| = |\text{a-ib}|.
\end{equation}
This is not the case if one uses the spatial operation of ``moment-reversal'' 
for the symmetry classification, 
since then both tensor elements $\chi^{(2)}_o$ and $\chi^{(2)}_e$ are just 
complex numbers without any constraints on their relative phase, see Fig. 
\ref{chi}b, and domain imaging is possible, as described in \cite{Artyk1}. 
Consequently, the symmetry analysis yields very different predictions if one 
uses time- or spin-reversal. In the limit far from resonances, however, the 
phase difference between $\chi^{(2)}_o$ and $\chi^{(2)}_e$ approaches 
$90^\circ$, and the domain contrast is lost also in the ``moment-reversal'' 
description (in agreement with experiment \cite{ref18}).

Finally we would like to remark on the validity of previous work on the 
group-theoretical classification of (magneto-)optical tensors. According to Pan 
{\em et. al.} \cite{ref53}, the time-reversal operation, because of its 
anti-unitarity, forces 
the tensor elements to decouple into mutually exclusive sets of purely real and 
imaginary ones (if all kinds of dissipation are neglected). In addition, the 
crystal symmetry 
forces the tensor elements to decouple into mutually exclusive sets of elements 
odd and even in {\em magnetization}-reversal, these two divisions are 
equivalent in the absence of conventional dissipation, i.e. real (imaginary) 
elements are even (odd) in the magnetization. These are the results of a purely 
quantum-mechanical approach, where the Hamiltonian is Hermitian 
(non-dissipative). However, the nonlinear susceptibility tensor describes the 
observed process of SHG, and thus one should not apply uniquely microscopic 
conclusions to the analysis of these tensor elements. Consequently, taking into 
account the dissipation in frequency space (i.e. redistribution of the response 
frequencies), will prevent the 
classification of tensor elements as purely real or imaginary ones, although for 
systems with higher symmetry the classification of tensor elements as odd and 
even ones in the magnetization (or in the antiferromagnetic order parameter {\bf 
L}) can still apply \cite{resonances}. 

In summary, we have shown that the time-reversal operation, often 
used for the symmetry classification of magneto-optical phenomena, in general 
cannot be applied to nonlinear optics. It should rather be replaced by spatial 
operations, resulting then in a proper description of the phenomena. 

The authors wish to thank Prof. P. Weinberger for pointing them to the concept 
of classical covering symmetries. We are also very grateful for interesting 
discussions with Dr. R. Vollmer. We acknowledge financial support by TMR Network 
NOMOKE contract no. FMRX-CT96-0015.

\end{multicols}

\begin{thebibliography}{99}
\bibitem{domains} This is possible e.g. in antiferromagnets like Cr$_2$O$_3$ or 
those with magnetic atoms of different kinds.
\bibitem{quantum} According to \cite{ref15}, both the $\sigma_x$ and $\sigma_y$ 
operations cause reversal of the spin part of the quantum-mechanical fermionic 
wavefunction. Of them, $\sigma_y$ is conveniently used to describe 
time-reversal, since it is an anti-unitary operation.
\bibitem{resonances} The nonlinear susceptibility tensor $\chi^{(2\omega)}$ was 
usually approximated to be real far from resonances. This approximation is not 
valid in the systems described by us: metals and transition metal oxides, where 
at any frequency one is close enough to one of the resonances (at least outside 
the gap of the latter). The crystals previously mostly used for SHG, and even 
more extensively as textbook examples, were usually wide-bandgap insulators and 
one is frequently far from any resonance.
\end{thebibliography}
\end{document}